\documentclass{aa}
\usepackage{natbib}
\usepackage{graphicx}
\usepackage{times}
\usepackage{amsmath}
\usepackage{txfonts}
\newcommand{\etal}{et al.,~}
\newcommand{\msun}{{\,\rm M}_{\odot}}
\newcommand{\lsun}{{\,\rm L}_{\odot}}

\newcommand{\kms}{\,{\rm km\,s}^{-1}}

\newcommand{\nht}{\ifmmode {{\rm NH}_3} \else {NH{\bas 3}} \fi}
\newcommand{\tco}{\ifmmode {^{13}{\rm CO}} \else {$^{13}{\rm CO}$}\fi}
\newcommand{\dco}{\ifmmode {^{12}{\rm CO}} \else {$^{12}{\rm CO}$}\fi}
\newcommand{\cdo}{\ifmmode {{\rm C}^{18}{\rm O}} \else {${\rm C}^{18}{\rm
O}$}\fi}

\newcommand{\htco}{\ifmmode {{\rm H}^{13}{\rm CO}^{+} } \else {${\rm H}^{13}
{\rm CO}^{+}$ }\fi}
\newcommand{\hco}{\ifmmode {{\rm H}^{12}{\rm CO}^{+} } \else {${\rm H}^{12}
{\rm CO}^{+}$ }\fi}
\newcommand{\ndhp}{\ifmmode {{\rm N}_{2}{\rm H}^{+} } \else {${\rm N}_{2}
{\rm H}^{+}$ }\fi}
\newcommand{\juz}{\ifmmode {{\rm J}=1\rightarrow 0} \else
{J=1$\rightarrow$0}\fi}
\newcommand{\jdu}{\ifmmode {{\rm J}=2\rightarrow 1} \else
{J=2$\rightarrow$1}\fi}
\newcommand{\jtd}{\ifmmode {{\rm J}=3\!\rightarrow\!2} \else
{${\rm J}=3\!\rightarrow\!2$} \fi}
\newcommand{\jcq}{\ifmmode {{\rm J}=5\!\rightarrow\!4} \else
{${\rm J}=5\!\rightarrow\!4$} \fi}
\newcommand{\as}{\ifmmode {^{\scriptscriptstyle\prime\prime}}
        \else $^{\scriptscriptstyle\prime\prime}$\fi}
\newcommand{\am}{\ifmmode {^{\scriptscriptstyle\prime}}
        \else $^{\scriptscriptstyle\prime}$\fi}
\begin{document}
\title{\bf Chemistry in disks \\
 I - Deep search for N$_2$H$^+$ in the protoplanetary disks around
LkCa~15, MWC~480, and DM~Tau.
\thanks{Based on observations carried out with the IRAM
Plateau de Bure Interferometer.
IRAM is supported by INSU/CNRS (France), MPG (Germany) and IGN
(Spain). Research partially supported by PCMI, the French national
program for the Physics and Chemistry of the Interstellar
Medium.}}

\author{Anne Dutrey \inst{1}, Thomas Henning \inst{2},
St\'ephane Guilloteau \inst{1}, Dmitry
Semenov \inst{2}, Vincent Pi\'etu \inst{3}, Katharina Schreyer
\inst{4}, Aurore Bacmann \inst{1}, Ralf Launhardt \inst{2},
Jer\^ome Pety \inst{3}, Frederic Gueth \inst{3}}
\offprints{Anne Dutrey, \email{anne.dutrey@obs.u-bordeaux1.fr}}
\institute{L3AB, Observatoire de Bordeaux, 2 rue de
l'Observatoire, BP 89, F-33270 Floirac, France \and
Max-Planck-Institut f\"ur Astronomie, K\"onigstuhl 17, D-69117
Heidelberg, Germany  \and IRAM, 300 rue de la piscine, F-38406
Saint Martin d'H\`eres, France \and Astrophysikalisches Institut
und Universit\"ats-Sternwarte, Schillerg\"asschen 2-3, D-07745
Jena, Germany}
\date{Apr 7, 2006 / Nov 21, 2006}
\abstract
{}
{To constrain the ionization fraction in protoplanetary disks, we present new high-sensitivity
interferometric observations of N$_2$H$^+$ in three disks surrounding DM~Tau, LkCa~15, and MWC~480.}
{We used the IRAM PdBI array to observe the N$_2$H$^+$ J=1-0 line and applied a $\chi^2$-minimization
technique to estimate corresponding column densities. These values are compared, together with HCO$^+$ column
densities, to results of a steady-state disk model with a vertical temperature gradient coupled to gas-grain
chemistry.}
{We report two \ndhp~detections for LkCa~15 and DM~Tau at the $5 \sigma$ level and an upper limit for
MWC~480. The column density derived from the data for LkCa~15 is much lower than previously reported. The
[N$_2$H$^+$/HCO$^+$] ratio is on the order of 0.02--0.03. So far, HCO$^+$ remains the most abundant observed
molecular ion in disks.}
{All the observed values generally agree with the modelled column densities of disks at an evolutionary stage
of a few million years (within the uncertainty limits), but the radial distribution of the molecules is not
reproduced well. The low inferred concentration of N$_2$H$^+$ in three disks around low-mass and
intermediate-mass young stars implies that this ion is not a sensitive tracer of the overall disk ionization
fraction.}
{}%
\authorrunning{Dutrey, Henning \etal}
\titlerunning{CID I: N$_2$H$^+$ in protoplanetary disks}
\maketitle
\section{Introduction}

A comprehensive understanding of the chemical composition and evolution of protoplanetary disks is a
necessary prerequisite to set up the physical and chemical conditions in which planet formation should occur.
Apart from CO and its isotopologues, and occasionally HCO$^+$, the molecular content of protoplanetary disks
remains poorly known. \citet{Dutrey_etal1997} first reported the discovery of a number of simple molecules
(HCN, CN, C$_2$H, H$_2$CO, and CS) in the circumstellar disk of DM~Tau and the circumbinary disk of GG Tau,
using the IRAM 30-m telescope, while \citet{Kastner_etal1997} observed HCN, HCO$^+$, and CO in TW Hydra.
Single-dish observations of molecular lines have also been reported for LkCa~15 \citep{Zadelhoff_etal2001,
Thi_etal2004} and for AB Auriga by \citet{Semenov_etal2005}.

More recently, millimeter arrays have started to search for molecular lines in the protoplanetary disks of
LkCa~15 \citep{Aikawa_etal2003,Qi_etal2003} and TW Hydra \citep{Qi_etal2004}. However, high-angular
resolution observations are extremely scarce and often noisy. Moreover, the simultaneous detection of the
continuum emission from the dust disk can bias the determination of the line intensity if not properly taken
into account.

Since molecular line data are limited in sensitivity and in spatial resolution, little is known about the
spatial distribution of molecules and their abundances in disks. As a consequence, comparison with existing
chemical models are hampered and often based on global data (e.g. the total number of molecules or, at best,
integrated line profiles). With these integrated quantities, it is difficult to test the validity of
sophisticated models, which include thermal structure and even time dependency.

Using the IRAM PdBI array, we started in 2004 the CID project ("Chemistry In Disks"), a molecular survey of
well-known disks such as those surrounding LkCa~15, MWC~480, and DM~Tau. Our main goal was to map
sufficiently strong molecular lines already detected in several disks with enough sensitivity to derive
molecular abundance variations versus radius. Among the molecules we searched for, molecular ions such as
HCO$^+$ or \ndhp are of great interest because they are potential tracers of the ionization structure of the
disks. Note that the latter molecular ion is usually used as a high-density tracer in dense molecular cloud
cores as it is supposed to barely freeze out onto dust grains even at low temperatures
\citep[e.g.][]{Crapsi_etal2005}.

In this first paper of a series we report the observations of \ndhp in three objects: DM~Tau, LkCa~15, and
MWC~480. Using the $\chi^2$-minimization technique in the $UV$ plane as described in
Guilloteau \& Dutrey \citep[1998; see also][~for details]{Pietu_etal2007}, we estimate the column density of
N$_2$H$^+$ in the outer parts of these disks and the corresponding uncertainties. We compare these observed
values with the N$_2$H$^+$ column densities computed with realistic 2-D steady-state disk models and a
gas-grain chemistry with surface reactions. We also briefly discuss the observed properties of HCO$^+$ in
these disks \citep[using results from][]{Pietu_etal2007} and deduce an estimate of the ionization fraction.

\begin{figure}[t]
\begin{center}
\resizebox{8.0cm}{!}{\includegraphics[angle=270]{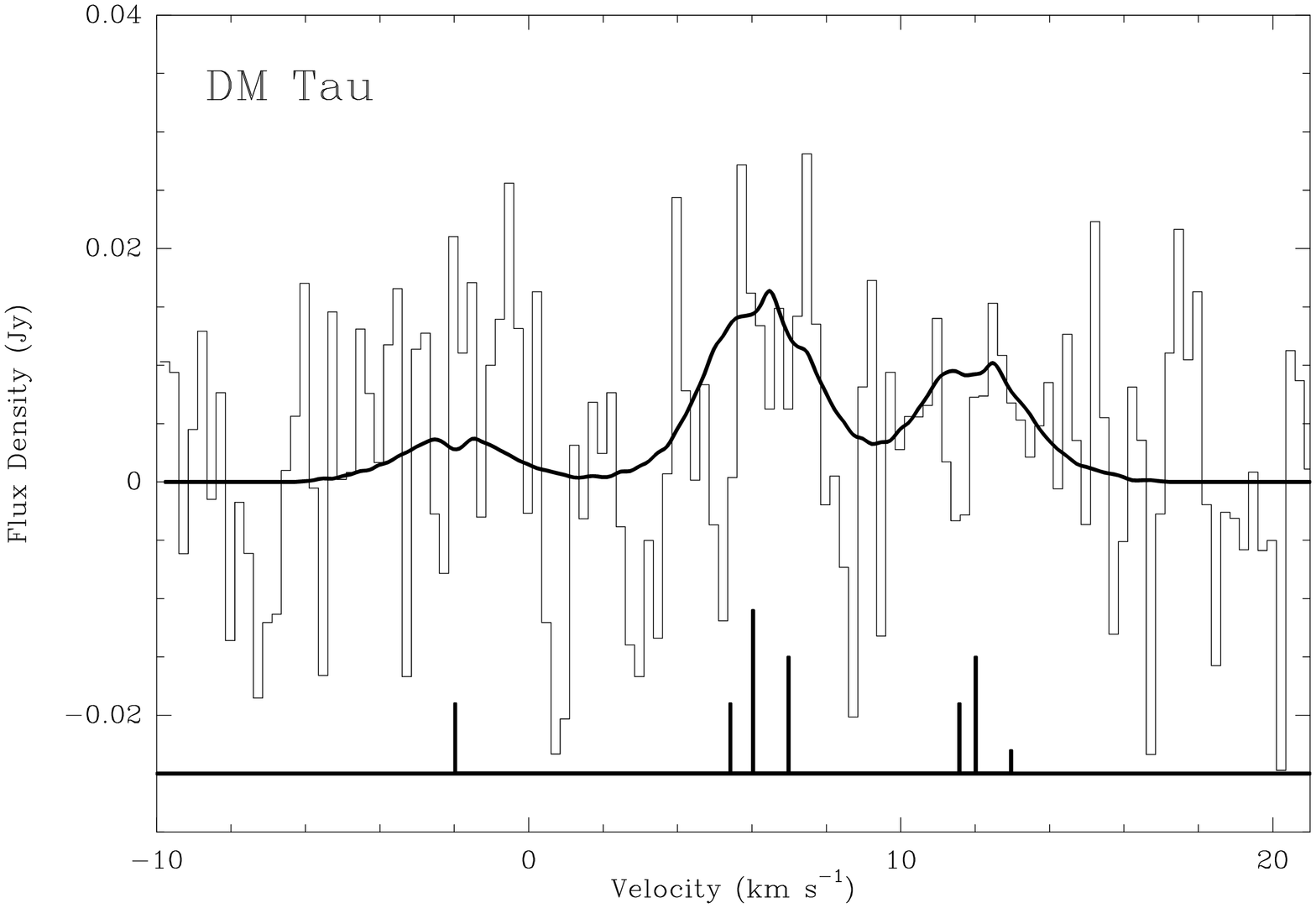}}
\resizebox{8.0cm}{!}{\includegraphics[angle=270]{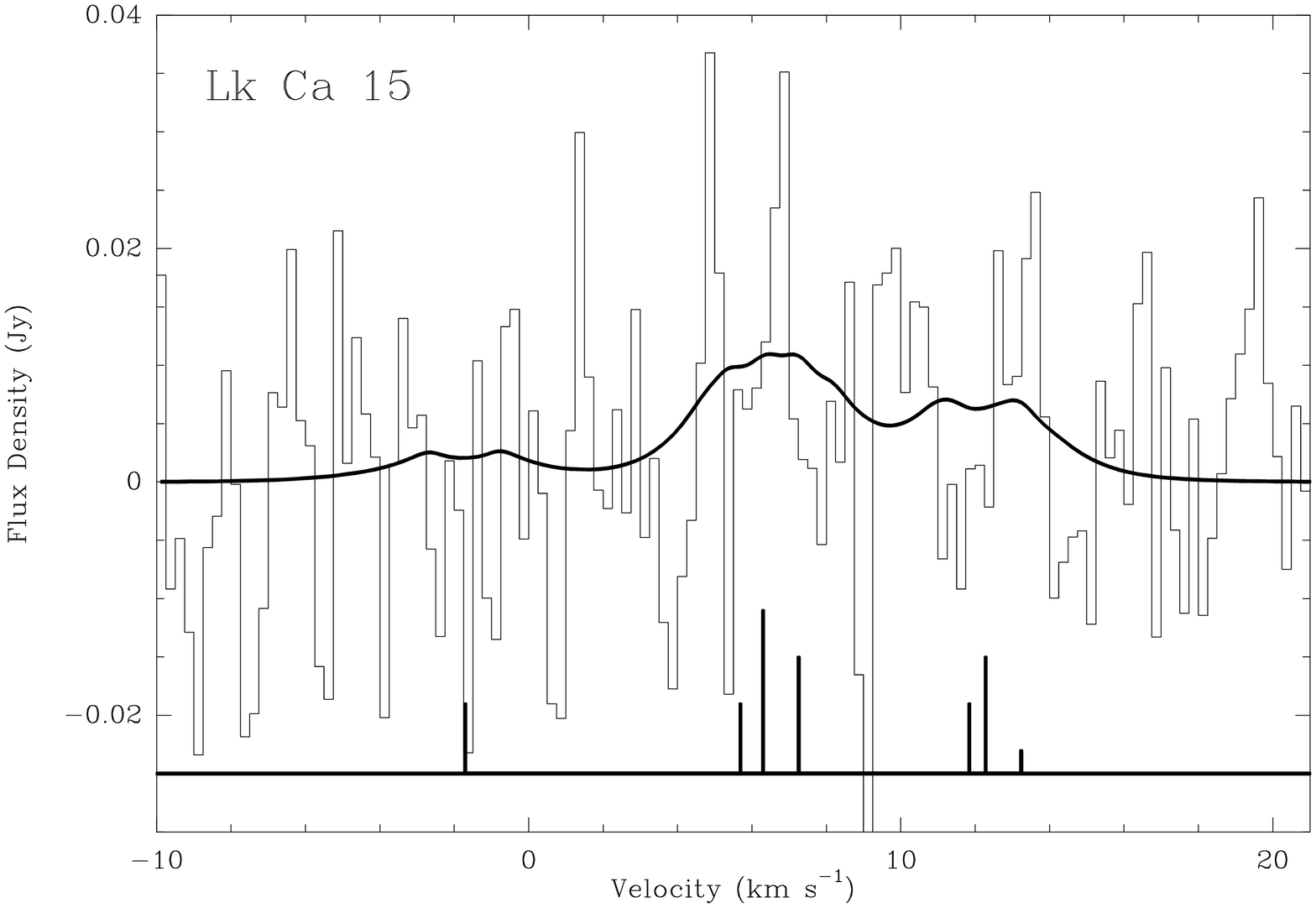}}
\resizebox{8.0cm}{!}{\includegraphics[angle=270]{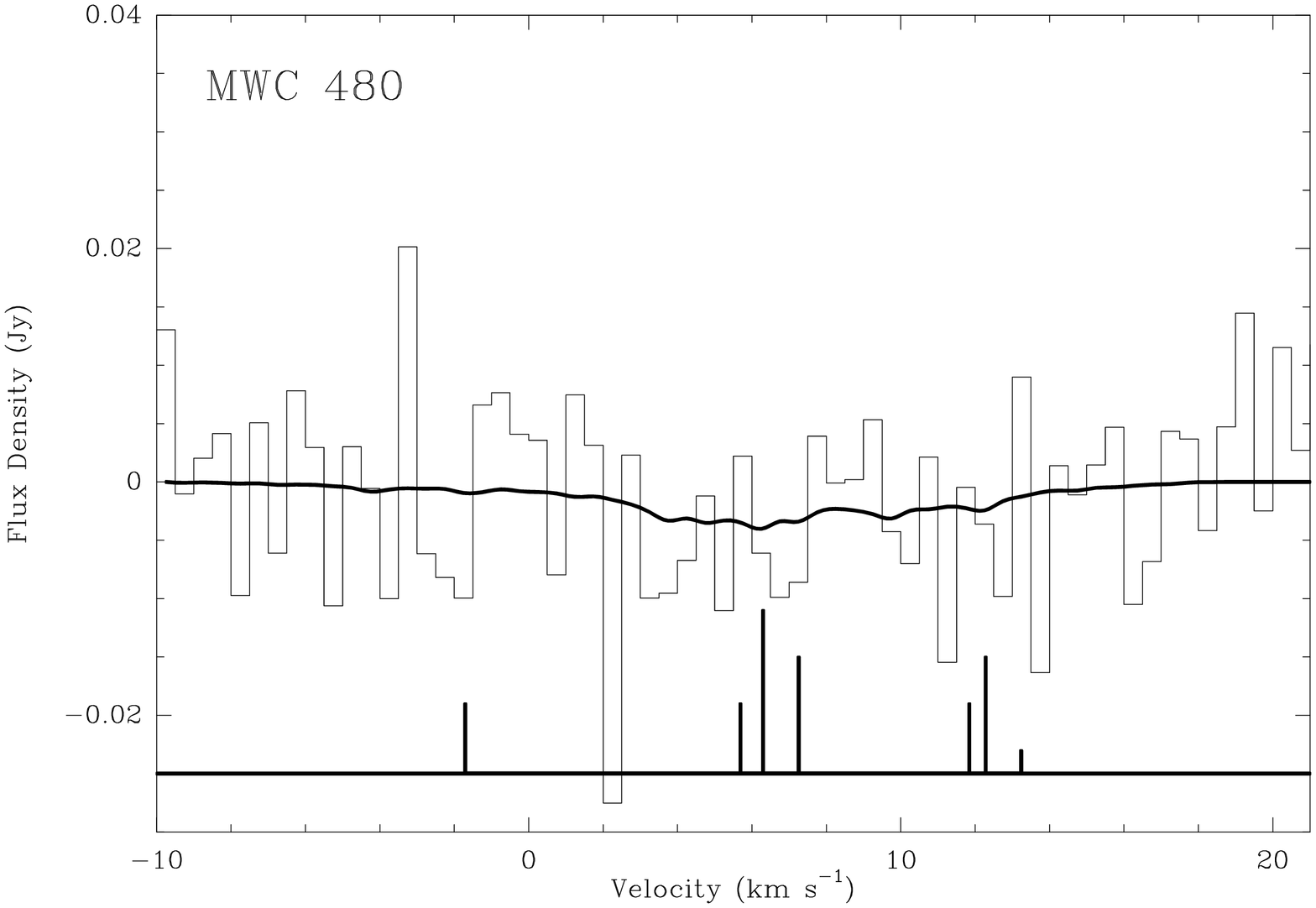}} \caption[Best models]{\label{fig:spectrum}
Integrated spectrum of \ndhp~\juz~ for DM~Tau (top), LkCa~15 (middle), and MWC~480 (bottom). The thick curve
represents the best model deduced from the $\chi^2$ minimization procedure. The location and relative
intensities of the hyperfine components are indicated for reference. The spectral resolution is 0.25 km
s$^{-1}$ for DM~Tau and LkCa~15 and is 0.5 km s$^{-1}$ for MWC~480. For DM~Tau and LkCa~15, the lines are
detected at the $5 \sigma$ level, but are difficult to identify visually because of the seven hyperfine
components.}
\end{center}
\end{figure}

In Sect.~2, we describe the observations and the sample of stars. The results are given in Sect.~3. Sect.~4
is dedicated to the corresponding chemical modelling. We discuss their implications in Sect.~5, followed by
conclusions.

\section{Sample of stars and observations}

\subsection{Sample of stars}

Table \ref{tab:coord} gives the coordinates and the physical properties of the observed stars. Our stars have
spectral types ranging from M1 to A4. We chose these objects because they have large dust and molecular
disks. Moreover, these stars are located in a hole or at an edge of their parent molecular cloud, in a region
devoid of CO emission.

\begin{table*}
\caption[]{Sample of Stars and Stellar Properties.}
\label{tab:coord}
\begin{center}
\begin{tabular}{lllllllll}
\hline \hline
Source   & Right Ascension         & Declination      & Spect.Type & Effective Temp. & Stellar Lum. & Stellar Mass & Age   & UV flux\\
         & ($^{o}$,$^{'}$,$^{''}$), (J2000.0) & ($^h$,$^m$,$^s$) (J2000.0) &            & (K)             & ($\lsun$)    & ($\msun$)    & (Myr) & ($G_0$) \\
\hline
LkCa~15  & 04:39:17.78 & 22:21:03.34  & K5 & 4350 & 0.74 & $1.01\pm0.02$ & 3-5 & 2550\\
DM~Tau   & 04:33:48.73 & 18:10:09.89  & M1 & 3720 & 0.25 & $0.53\pm0.03$ & 5   &  410\\
MWC~480  & 04:58:46.26 & 29:50:36.87  & A4 & 8460 & 11.5 & $1.83\pm0.05$ & 7   & 8500\\
 \hline
\end{tabular}
\end{center}
Coordinates J~2000.0 deduced from the fit of the 1.3mm continuum map of the PdBI performed in
\citet{Pietu_etal2007}. Col.3,4, 5, and 7 are the spectral type, effective temperature, the stellar
luminosity and age as given in \citet{Simon_etal2000}. Masses are as given by \citet{Pietu_etal2007}. The
stellar UV fluxes that are given in Col. 8 in units of the \citet{Draine1978} interstellar UV field are taken
from \citet{Bergin_etal2004} (LkCa~15 and DM~Tau) or computed from the \citet{Kurucz1993} ATLAS9 of stellar
spectra (MWC~480). They are given for a distance of 100 AU from the star.
\end{table*}

\begin{figure*}[t]
\begin{center}
\resizebox{14.0cm}{!}{\includegraphics[angle=270]{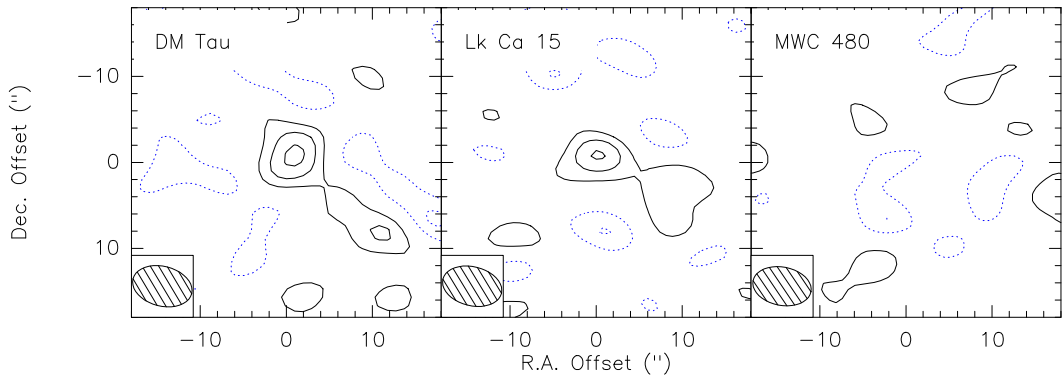}} \caption[data]{\label{fig:maps} Weighted
intensity maps (see text for details) of \ndhp \juz~ towards DM~Tau, LkCa~15, and MWC~480. The
interferometric beam is $7.0'' \times 4.6''$, with P.A. 75$^\circ$ or 1000 AU $\times$ 650 AU at the Taurus
distance. Contours are approximately 1.5 $\sigma$.}
\end{center}
\end{figure*}

\subsection{Observations}

Observations of \ndhp were carried out in November 2004 and in April and September 2005. The observing time
was shared in each track between LkCa~15, MWC~480, and DM~Tau. The D configuration was used with 6 antennas.
The sources were observed in time-sharing mode during 3 transits, leading to a  total on source integration
time of about 6 hours for each object. At 3\,mm, the tuning was single-side-band (SSB). The backend was a
correlator with three bands of 20 MHz (spectral resolution 0.125 $\kms$) centered on the \ndhp lines, one
band of 20 MHz centered on the CS J=5-4 line, and 2 bands of 320 MHz for the 1.3mm and 3mm continuum,
respectively. The phase and flux calibrators were 0415+379 and 0528+134. Observations of HCO$^+$ are reported
in \citet{Pietu_etal2007}. At 3\,mm, the primary beam of the 15-m antennas is on the order of 56$''$, or
$\sim 8000$~AU at the Taurus distance.

For both data sets, we took great care with the relative flux calibration by using not only MWC~349 as a
common reference (see the IRAM flux report, number 11) but also by always having our own internal reference
in the observations. For this purpose we applied two methods: i) we used MWC~480 as a reference because its
continuum emission is reasonably compact and bright; ii) we also checked the consistency of the flux
calibration using the spectral index of DM~Tau, as described in \citet{Dartois_etal2003}. We cross-checked
both methods and thus obtained a reliable relative flux calibration from one frequency to another. Using our
method, we estimated the relative accuracy of our flux density scale to be significantly better than $10 \%$
at 3\,mm.

\section{Results}

Integrated spectra (over a $6'' \times 6''$ region) are displayed in Fig.\ref{fig:spectrum}, superimposed
with the best-fit disk model, as derived from the disk modelling described below.

The \ndhp~\juz~ emission is very weak, and determination of the integrated area is complicated by the
hyperfine structure. Furthermore, due to the Keplerian velocity gradient, some components overlap in
velocity. Half of the signal is in the satellite components: computing the integrated line flux would require
us to include these low S/N data using twice more bandwidth than for the main group. This would degrade the
overall S/N ratio. To better illustrate the detection, we instead applied a filter in which each channel is
weighted by the expected line intensity (obtained from the best fit), and the sum over channels was
normalized by the sum of weights, a classical optimal filtering. This method confirms the detection but does
not provide a quantitative estimate, which is given by the minimization (see below). The resulting weighted
intensity maps are displayed in Fig.\ref{fig:maps}. Weak but significant detections are visible at the
position of DM~Tau and LkCa~15, but no signal is detected towards MWC~480.

The column density of N$_2$H$^+$ is not a directly observable quantity, but it must be deduced from the map
intensities, taking the line formation mechanism into account. In the analysis of disk line data the surface
density is often assumed to be a power law, and then the power law index is fixed during the analysis. In
this paper we follow a different approach by applying a minimization technique that results both in the
N$_2$H$^+$ column density and the power law index.

Since the analysis of the \ndhp \juz~line is complicated by its hyperfine structure, we decided to estimate
the \ndhp column density by directly comparing the interferometric data to molecular disk models. Following
\citet{Guilloteau_Dutrey1998}, the data were analyzed inside the $UV$ plane. For each source, we fit a
passive Keplerian disk model that assumes power-law radial dependencies for the molecular rotation
temperature ($T_r(r)= T_{100}\times (\frac{r}{100\rm{AU}})^{-q}$) and the surface density ($N_{\ndhp}(r) =
N_{100} \times (\frac{r}{100\rm{AU}})^{-p}$). We assume hydrostatic equilibrium with a scale height derived
from the CO measurements \citep[see][]{Pietu_etal2007}, and a fixed local line-width $\Delta v$. The model
also takes into account the stellar mass (Table \ref{tab:coord}) and disk inclination (Table
\ref{tab:disks}).
The disk parameters were taken as constant during the $\chi^2$ minimization process.  Only the
surface-density law parameters ($N_{100}$ and $p$) of \ndhp were varied during the minimization process. For
thermalized lines such as CO J=1-0 and J=2-1 or HCO$^+$ J=1-0, the rotation temperature is a direct estimate
of the gas kinetic temperature. In our case, we take the rotation temperature law from HCO$^+$ data
\citep[see Table \ref{tab:fit} and][for details]{Pietu_etal2007}. Such a low temperature implies that the
HCO$^+$ J=1-0 emission is coming from the mid-plane. Given the expected temperature range that is measured in
these disks (see Tables \ref{tab:disks}-\ref{tab:fit}), the chosen values have little influence on the
derived column density.

\begin{figure}[t]
\begin{center}
\resizebox{8.0cm}{!}{\includegraphics[angle=270]{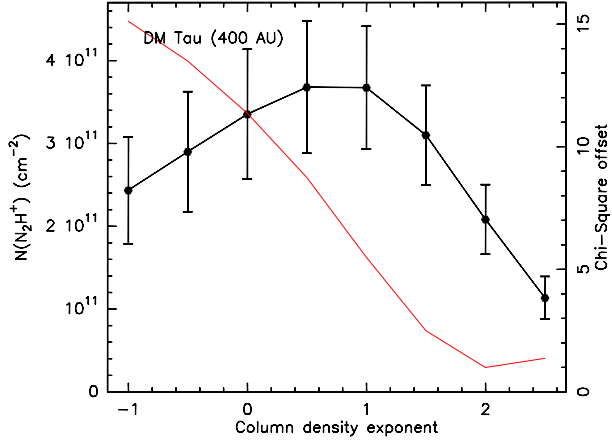}} \caption[Best fit for
DM~Tau]{\label{fig:fitdmtau} Surface density of \ndhp\ at 400 AU (with error bars) versus radial exponent for
DM~Tau, the error bars are for $\pm 1$ formal $\sigma$. The units are specified on the left Y axis. The light
grey curve is the $\chi^2$ (with an offset), with units on the right Y axis.}
\end{center}
\end{figure}

\begin{figure}[t]
\begin{center}
\resizebox{8.0cm}{!}{\includegraphics[angle=270]{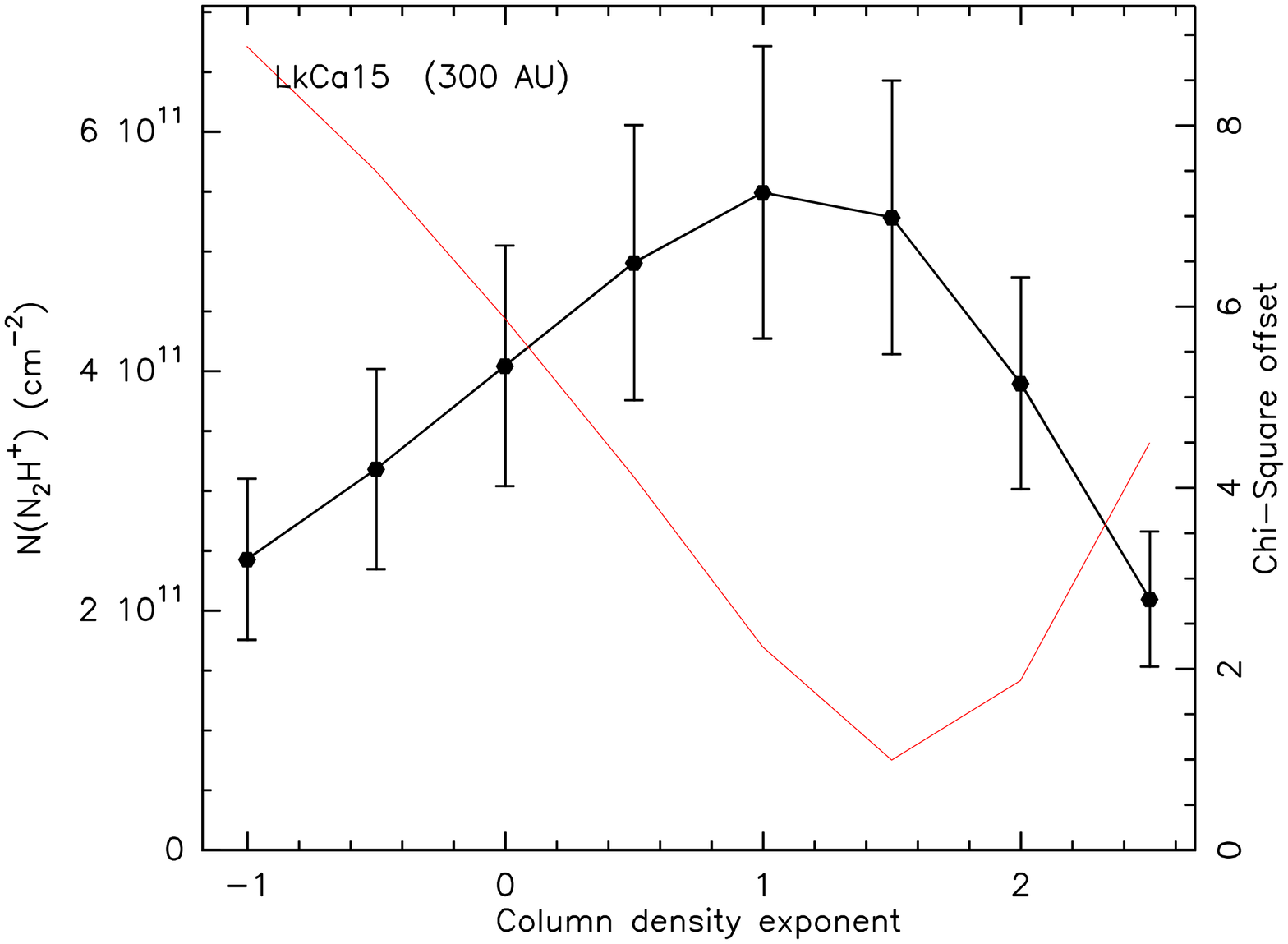}} \caption[Best fit for
LkCa~15]{\label{fig:fitLkCa15} Surface density of \ndhp\ at 300 AU (with error bars) versus radial exponent
$p$ for LkCa~15, the errors bars are for $\pm 1$ formal $\sigma$. The units are specified on the left Y axis.
The light grey curve is the $\chi^2$ (with an offset), with units on the right Y axis.}
\end{center}
\end{figure}

The results of the fits are summarized in Table \ref{tab:fit} and presented in Fig.~\ref{fig:fitdmtau} for
DM~Tau and in Fig.~\ref{fig:fitLkCa15} for LkCa~15, respectively. We have chosen to show the results at
400~AU for DM~Tau (300~AU for LkCa15) because it is the relevant scale for the IRAM data in terms of
sensitivity and resolution. These figures indicate that we detected \ndhp\ at the 4 -- 5 $\sigma$ level in
both sources, with a best-fit exponent of 2 -- 2.5. For MWC~480, the formal best fit leads to negative column
densities, at the  $1 \sigma$ level: we thus only mention a $3 \sigma$ upper limit for this source.

\begin{table}
\caption{Disk Parameters} \label{tab:disks}
\begin{center}
\begin{minipage}[t]{\columnwidth} \centering
\renewcommand{\footnoterule}{}  
\begin{tabular}{lccccc}
\hline \hline Source   & $i$    & R$_{out}$ & $T_{100}$ & Acc.Rate & Disk Mass \\
                & ($^o$) & (AU)      & (K) & ($\msun$\,yr$^{-1}$) & ($\msun$) \\
\hline
DM~Tau   & -32 & 800 & 15,   q$\sim 0.5$ & 2.10$^{-9}$ & 0.05 \\
LkCa~15  &  52 & 680  & 15,   q$\sim 0.5$ & 10$^{-8}$   & 0.03 \\
MWC~480  &  38 & 695\footnote{This value is taken from \citet{Thi_etal2004}.} & 30,   q$\sim 0.5$ & 10$^{-8}$   & 0.03 \\
\hline
\end{tabular}
\end{minipage}
\end{center}
Col.2 Inclinations are taken from \citet{Simon_etal2000}.~ Col.3 \& 4 Outer radius and temperature are taken
from the \tco~ interferometric analysis of \citet{Pietu_etal2007}. Note that the temperatures given here
correspond to the observed (CO data) gas kinetic temperature from the cold mid-plane region and is higher in
the disk intermediate layer. They are in good agreement with predictions from the model by D'Alessio and
collaborators. Col.5 Accretion rates are taken as a generic value from \citet{DAlessio_etal1999} for MWC~480
and LkCa~15. The accretion rate for DM~Tau is taken from \citet{Calvet_etal2005}. Col.6 Disk masses are from
\citet{Dutrey_etal1997} for DM~Tau, \citet{Thi_etal2004} for LkCa~15 and \citet{Mannings_etal1997} for
MWC~480.
\end{table}
\begin{table}\caption{Column densities of HCO$^+$ and N$_2$H$^+$ and abundance ratios}
\label{tab:fit}
\begin{center}
\begin{tabular}{lccc}
\hline\hline
Source        & MWC~480         &  LkCa~15 & DM~Tau \\
\hline
T  & 20  & 19   & 13   \\
q & 0.4 & 0.38 & 0.15 \\
\hline
$N$(HCO$^+$) & $  5.0 \pm 0.6 \,10^{12} $ & $ 8.0 \pm 0.5 \,10^{12} $ & $ 6.5  \pm 1.4 \,10^{12} $ \\
$p$       & $   2.1 \pm 0.5 $ & $ 2.5 \pm 0.3 $ & $ 2.7 \pm 1.0 $ \\
R       &   250  AU &  300 AU & 400 AU \\
\hline
$N$(N$_2$H$^+$) & $ < 3 \,10^{11} $ & $ 2.3 \pm 0.5 \,10^{11} $ & $ 1.1 \pm 0.25\,10^{11} $ \\
$p$    & $ 2  $ & $ 2.5 $ & $ 2.5 $ \\
R &  250 AU   &  300 AU  &  400 AU \\
\hline
{\scriptsize [N$_2$H$^+$]/[HCO$^+$]} & $ < 0.06 $ & $\sim 0.03$ & $\sim 0.02$ \\
\end{tabular}
\end{center} $T$ indicates the temperature (in K) at 100 AU  and $q$ the temperature exponent used to derive
the column densities, taken from \citet{Pietu_etal2007}. For DM~Tau and LkCa~15, in agreement with the
figures, we present the results for $p=2.5$. Columns densities are in cm$^{-2}$. Values for HCO$^+$ are taken
from \citet{Pietu_etal2007}.
\end{table}

The $5 \sigma$ detection in LkCa~15 corresponds to a column density that is apparently about a factor $\sim$
100 below the column density deduced from OVRO data by \citet{Qi_etal2003}, who report an observed value of
$1.7 \, 10^{13}$ cm$^{-2}$. The difference can be traced down to several factors. \citet{Qi_etal2003}
indicate a total line flux of 0.3 Jy km/s (from their Table 1), while our best fit line flux is 0.15 Jy km/s.
Furthermore, \citet{Qi_etal2003} use a constant temperature of 30 K, and give the beam-averaged column
density ($p=0)$, which corresponds to a source radius of $\sim$ 280 AU. Under the same conditions, we derive
a column density of $4.5 \pm 0.9~10^{12}$ cm$^{-2}$. Assuming that the \citet{Qi_etal2003} flux corresponds
to the main group of hyperfine components resolves the discrepancy and shows we have detected 4 times less
N$_2$H$^+$ molecules than claimed by \citet{Qi_etal2003} if we take the same conditions for the analysis.
Part of the difference may be due to the treatment of the continuum: when computing the integrated emission,
we subtracted the continuum while \citet{Qi_etal2003} included it with the justification that the lines are
optically thick. The assumption that the lines are optically thick clearly does not agree with intensities
detected so far in disks, as HCO$^+$(1-0) \citep{Semenov_etal2005,Pietu_etal2007}, C$^{18}$O(1-0) and
$^{13}$CO(1-0) \citep{Dartois_etal2003}, and N$_{2}$H$^{+}$(1-0) are optically thin. Furthermore, even for an
optically thick line the continuum should be subtracted because at a given velocity in the disk, the line
emission covers only a tiny fraction of the entire continuum, which is at most $dV/ V_{\rm Kepler}$ where
$dV$ is the local line-width and $V_{\rm Kepler}$ the projected rotation velocity at the disk edge
\citep{Beckwith_Sargent_1993, Guilloteau_etal2006}. Thus the integrated area of the line will be
significantly overestimated if the continuum is not subtracted.

The large discrepancy between our analysis and that of \citet{Qi_etal2003} clearly demonstrates the necessity
of taking proper care of the column density distribution. When comparing molecules with different apparent
extents or observed with different interferometric beams, this is fundamental in order to avoid biases due to
sensitivity limitations. In our case, we found $p=2.2\pm0.8$ for DM~Tau and $p=1.6 \pm0.6$ for LkCa~15. These
values are deduced from the best \ndhp models and are in complete agreement with the analysis performed for
the most sensitive tracer HCO$^+$. Most importantly, we derive [N$_2$H$^+$]/[HCO$^+$] under the assumption
that this ratio is constant with radius. This is justified both by the above agreement between the $p$ values
and by the chemical models presented in Sect.4.

This result should serve as a warning that ``observed'' column densities either (i) come with assumptions for
the surface density profile or (ii) are based on a fitting procedure within physical models. In our work, the
terms ``observed column densities'' refer to values derived using the second method. The expression
``modelled column densities'' refers to column densities predicted by a chemical model, such as the one
described in the next section.

\section{Chemical modelling}
\label{chem_model}

Since our interferometric millimeter data are sensitivity limited, we have chosen to focus on the more
prominent chemical effects. In this paper, our main goal is to reproduce the overall distribution of the
observed molecular ions. Accordingly, the chemical model we use does not take into account grain growth and
sedimentation. Similarly, \citet{Semenov_etal2006} have shown that turbulent transport can modify the column
densities and abundance distributions of some molecules with respect to the usual static chemical models in a
number of cases. However, we do not include vertical/radial mixing here.

\subsection{Model description}

In order to confront the observed values with theoretical predictions, we used an updated gas-grain chemical
model with surface reactions from \citet{Semenov_etal2005}. Briefly, it is based on the UMIST'95 gas-phase
network \citep{Millar_etal1997} supplemented with surface chemistry on uniform 0.1$\mu$m silicate grains from
\citet{Hasegawa_etal1992} and \citet{Hasegawa_etal1993} and a small set of deuterium chemistry reactions
(E.A.~Bergin, private communication). Using measurements published during the last decade, we updated a few
tens of important reaction rates in the UMIST'95 database \citep{Vasyunin_etal2006}. Overall, our chemical
network consists of 560 species made of 14 elements and about 5400 reactions.

The primary ionization sources in this model include stellar X-ray and UV radiation, interstellar UV
radiation, and cosmic ray particles, and the decay of short-lived radionuclides. The adopted intensity of the
un-attenuated stellar UV flux at 100~AU from the central stars are given in the last column of
Table~\ref{tab:coord}. We used the standard interstellar UV field of \citet{Draine1978} and re-scaled it to
match the corresponding stellar luminosities of TTauri stars according to the current limited knowledge of
their UV spectra. This approach should be improved as soon as better UV data become available
\citep[see][]{Bergin_etal2006}. Note that the UV penetration is computed using a 1-D plane-parallel approach
and thus the UV intensity in a given disk location is very likely underestimated in comparison with the full
2-D radiative transfer simulations; see \citet{vanZadelhoff_etal2003}. The intensity and penetration of the
stellar X-ray radiation are modelled using observational results of \citet{Glassgold_etal2005} and
Monte-Carlo simulations of \citet{GNI97a,GNI97b}. The corresponding un-attenuated ionization rates at 100~AU
are equal to $4.5\,10^{-14}$~s$^{-1}$ for DM~Tau, $2\,10^{-13}$~s$^{-1}$ for LkCa~15, and
$2\,10^{-15}$~s$^{-1}$ for MWC~480.

Using the results of laboratory measurements by \citet[ their ``mixed-ice'' case]{Bisschop_etal2006}, we
assumed that desorption energies of CO and N$_2$ are equal to 930~K. For all sources, we used a flared 1+1D
disk model with a vertical temperature gradient similar to that of \citet{DAlessio_etal1999}. This disk model
has the $\sim 30$~K temperature surface at $\sim 3$ pressure scale heights (the CO surface where $\tau \sim
1$ for the CO J=2-1 transition) at a distance of 100~AU, in agreement with the results obtained by
\citet{Dartois_etal2003} and \citet{Pietu_etal2007}. Other disk parameters are summarized in
Table~\ref{tab:disks}.

With this physical and chemical model in hand and TMC1-like initial abundances, the chemical evolution is
simulated over the 5~Myr (DM~Tau and LkCa~15) and 7~Myr (MWC~480) evolutionary time. We repeated the
modelling of the disk chemical evolution 20 times by using the same chemical network and varying its rate
coefficients within their uncertainty limits, similar to \citet{Vasyunin_etal2004} and
\citet{Wakelam_etal2005}. This allows us to get first-order estimates of the intrinsic uncertainties of the
theoretical abundances.

\subsection{Ionization in disks and global evolution of N$_2$H$^+$ and HCO$^+$}

\begin{figure*}
\centering \includegraphics[width=13.0cm]{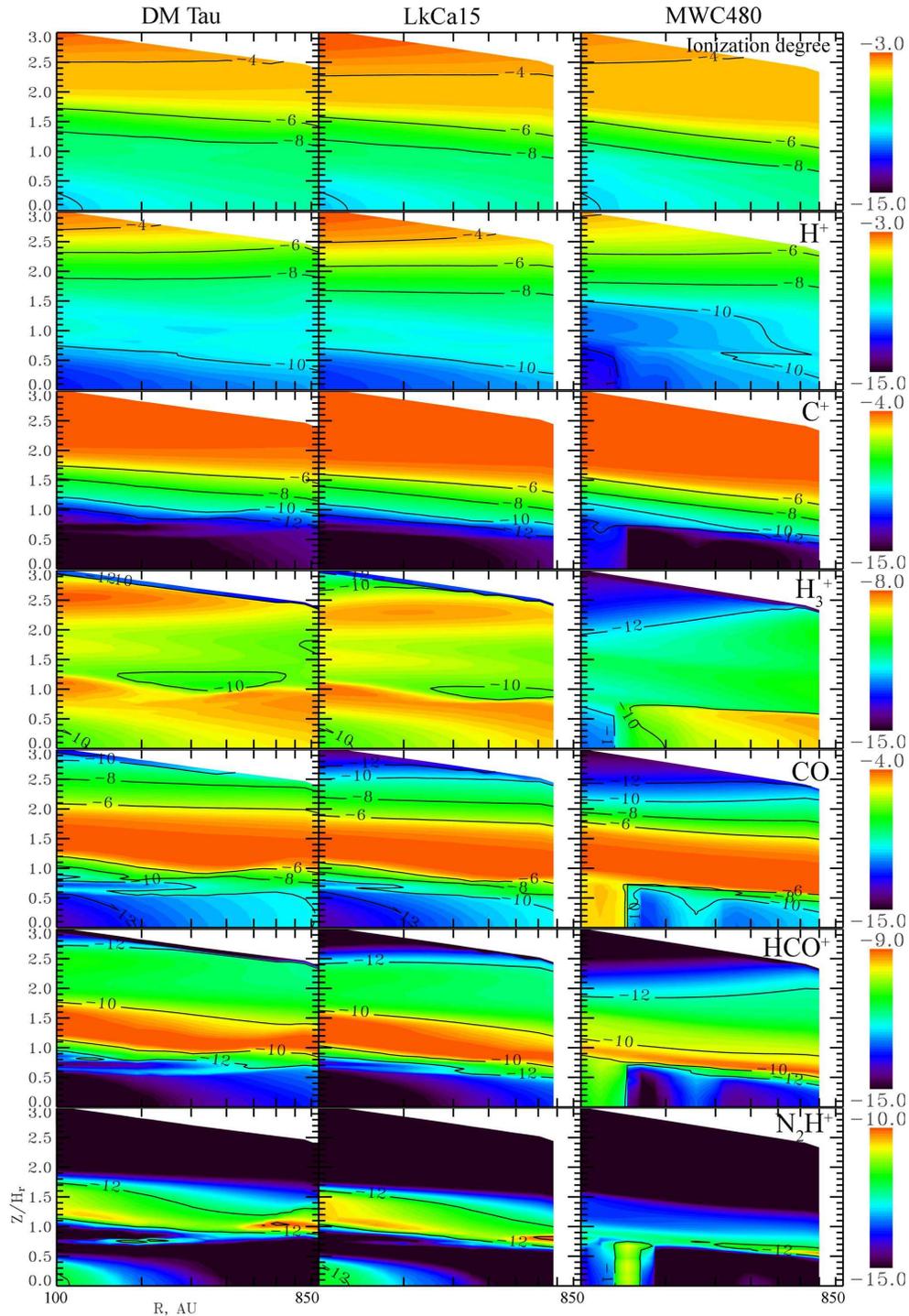} 
\caption[All abundance distributions in the disks after a few Myr of evolution] {From top to the bottom:
model-averaged relative abundance distributions (in respect to the total number of hydrogen nuclei) of
important ions: H$^+$, C$^+$, H$_3^+$, HCO$^+$, N$_2$H$^+$, and CO in the DM~Tau (left), LkCa~15 (middle),
and MWC~480 (right panel) disks after 5 and 7 Myr of the evolution. The Y-axis is the disk vertical extent
normalized by one pressure scale height (different for all three objects).} \label{fig:all_abunds}
\end{figure*}

In Fig.~\ref{fig:all_abunds} we present the model-averaged disk abundances of CO, several dominant ions, and
the ionization fraction after a few Myr of evolution for all three objects without mixing. The Y-axis is
expressed in units of the pressure scale height, $H_r$, as defined in \citet{Dartois_etal2003}.

More generally, in the upper surface layer ($\ga 2$ scale heights) the dominant ion is H$^+$, with an
abundance up to $\sim 10^{-3}$ (with respect to the total amount of hydrogen nuclei). Somewhat lower, at $\ga
1.5$ scale heights, the disk ionization degree is determined by ionized carbon, which is also the most
abundant ion in all three disks ([C$^+$] $\la 10^{-4}$, $N({\rm C}^+) \approx N({\rm e}^-) \sim 3\,
10^{16}$~cm$^{-2}$ around $\sim$ 200-500 AU). The evolution of both ions is governed entirely by fast
photo-reactions due to energetic stellar UV field and X-rays and interstellar UV radiation. In contrast, the
less abundant H$_3^+$ ion mostly traces the fractional ionization closer to the disk midplane, [H$_3^+$]
$\sim$ [e$^-$] $\la 10^{-9}$, where the cosmic ray particles remain the only ionization source.
Unfortunately, all these important ions are not easily detectable in protoplanetary disks \citep[however,
see][]{Kamp_etal2003,Ceccarelli_Dominik2005}.

The \emph{observationally} most important ion in the outer disks is predicted to be HCO$^+$ (typical
abundance value is $\la 10^{-9}$). It is located in the warm molecular layer that is partly shielded from
direct ionizing radiation from the central star and the ISM, at $\sim 1$ pressure scale height. Note that the
corresponding HCO$^+$ layer is thinner and located deeper towards the midplane in the MWC~480 disk in
comparison with DM~Tau and LkCa~15. Such a chemical effect is caused primarily  by the strong radiation field
of MWC~480, which is capable of dissociating CO -- the key molecule for the HCO$^+$ evolution -- more deeply
inside the disk, as seen in the figure. The more intense stellar heating also leads to a warmer interior of
the MWC~480 disk compared to the cases of DM~Tau and LkCa~15, which in turn results in efficient CO
desorption in the MWC~480 disk midplane at $R\la200$~AU and thus in a reservoir of abundant HCO$^+$ there.
Note that, in all three cases, the upper boundary of the abundant HCO$^+$ layer coincides with the lower
boundary of the C$^+$ layer that marks the disk height where penetration of the ionizing radiation
essentially stops. This happens at $\sim 1-1.5$ scale heights.

In contrast to HCO$^+$, the next observationally important ion N$_2$H$^+$ is concentrated in a very thin
layer at the lower bottom of the intermediate layer ($\sim 0.5-1$ scale heights) and in the disk midplane.
Unlike for other considered species, \ndhp abundances are more spatially peaked toward outer disk regions. It
is also interesting that the \ndhp vertical layer is in general thinner in the MWC~480 disk compared to the
disks around LkCa~15 and DM~Tau, similar to the HCO$^+$ distribution. The main reason for the chemical
differentiation between these two ions is the fact that \ndhp gets  destroyed easily in proton transfer
reactions with CO, e.g. \ndhp + CO $\rightarrow$ HCO$^+$ + N$_2$
   \citep[$k=8.8\,10^{-10}$~cm$^3$\,s$^{-1}$][]{Anicich_Huntress86} .

Thus, \ndhp is less abundant in the disk region where the CO concentration is high, that is, in the CO
molecular layer above about one pressure scale height. On the other hand, it also cannot be abundant in the
disk surface layer exposed to the energetic radiation from the star, which destroys all polyatomic species
within $\sim 10^4$ years. Finally, it cannot reach high levels of concentration in the outer disk midplane
($R>400$~AU in DM~Tau and LkCa~15, and $R>200$~AU in MWC~480) because the evolution of \ndhp in the entire
midplane region is governed by a cycle between ion-molecule reactions of N$_2$ with the most abundant ion,
H$_3^+$ ($\la 10^{-9}$), and dissociative recombination of \ndhp with electrons and charged grains.
Consequently, at very low temperatures in the outer midplane a volatile product of the \ndhp dissociation,
namely NH (desorption energy is 600~K), is frozen out on grain surfaces and eventually converted to solid
NH$_3$ by a two-step process of surface hydrogenation:
\begin{displaymath} \mbox{NH + H $\overrightarrow{grains}$ NH$_2$ + H
$\overrightarrow{grains}$ NH$_3$}~. \end{displaymath} Thus, after about 1~Myr of evolution all nitrogen is
locked in solid ammonia that has an abundance of about $10^{-5}$. In contrast, in the warmer disk midplane
region that is located closer to the central star, NH is not totally frozen out and thus the surface
formation of ammonia is not that efficient. Therefore, about 10\% of all nitrogen is locked in solid N$_2$
($\sim 2\, 10^{-6}$), but a minor fraction of molecular nitrogen remains in the gas phase ($\sim 10^{-8}$)
and reacts with H$_3^+$, forming N$_2$H$^+$.

Note that there is a thin layer devoid of N$_2$H$^+$ at about 0.7 scale height in the disk of DM~Tau at
100~AU  and, to a lesser extent, LkCa~15. In contrast to the Herbig Ae star MWC~480, these two Sun-like T Tau
stars are strong emitters of hard X-ray radiation \citep{Glassgold_etal2005}. The stellar X-ray radiation
cannot propagate directly through the disk midplane, but in general has larger penetration depth toward the
disk interior than UV photons. \citet{Semenov_etal2004} have shown that the attenuated X-rays can drive a
peculiar chemistry leading to the production of complex (organic) species on dust grains.

\begin{figure*}
\begin{center}
\resizebox{0.9\hsize}{!}{\includegraphics[angle=270]{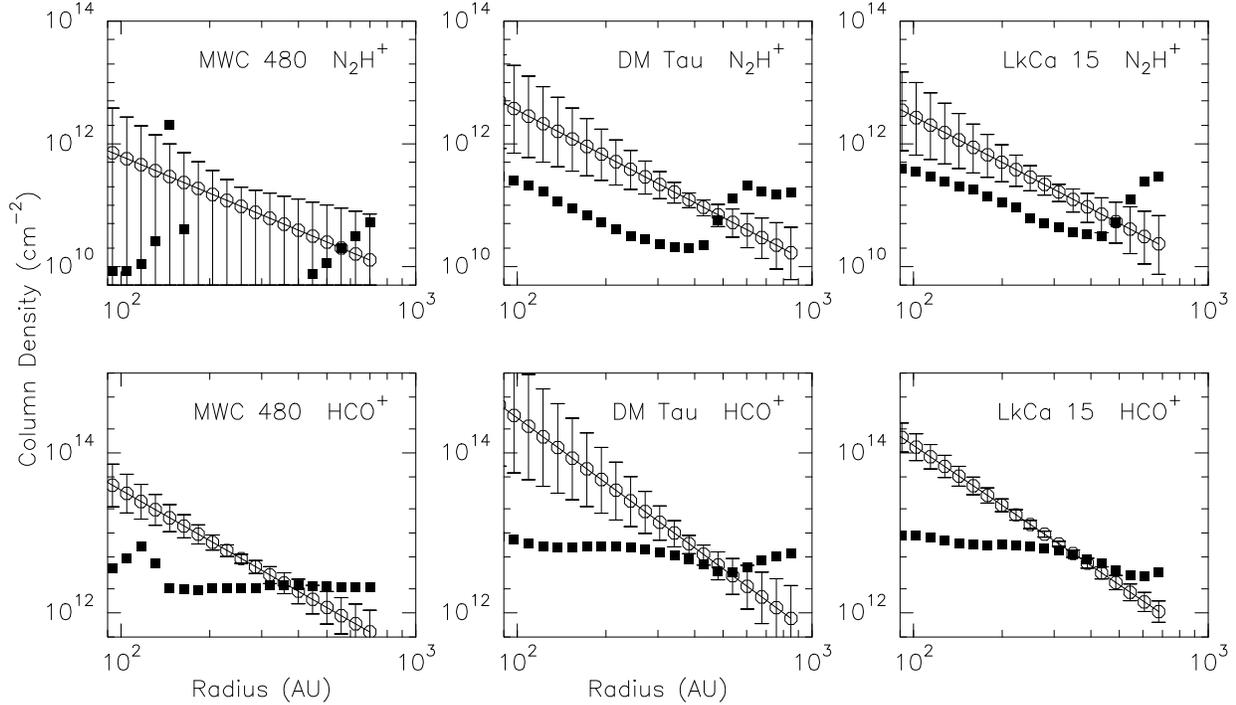}} \caption[Column densities versus radius
after a few Myr of evolution]{\label{fig:coldens} Top: Expected column density of \ndhp\ versus radius (black
dots), compared to the observed column densities (curve and errorbars) in MWC~480 (left), DM~Tau (middle),
and LkCa~15 (right). Bottom: as above but for HCO$^+$. The disk age is 5 Myr for DM~Tau and LkCa~15 and 7 Myr
for MWC~480. The abrupt changes in predicted column densities of \ndhp at some radii are partly due to the
limited resolution of the chemical model (see Fig. \ref{fig:all_abunds}). For HCO$^+$, the error bars include
the (measured) uncertainty on the excitation temperature for DM Tau, while this excitation temperature was
imposed for LkCa~15 and MWC~480}
\end{center}
\end{figure*}


\section{Discussion}

\subsection{Observed abundances of molecular ions in disk}

HCO$^+$ is observed in many astrophysical media with a wide variety of physical conditions, from diffuse
clouds \citep{Liszt_etal2001} up to dense cores \citep{Caselli_etal2002a}. N$_2$H$^+$, which can be
significantly less abundant and even depleted in very dense cores \citep{Pagani_etal2005}, appears more
difficult to observe \citep{Tafalla_etal2004}. Moreover, as described in Sect.~4.2, \ndhp~can be destroyed by
reactions with CO to give HCO$^+$ \citep[see also,][]{Anicich_Huntress86}. Hence, HCO$^+$ and \ndhp usually
do not coexist. Having abundance estimates coming from the same area is not so easy. This is, for example,
illustrated by \citet{Jorgensen_2004} in L483-mm where he finds that the HCO$^+$ traces the large scale
outflow, while \ndhp~traces the quiescent material (in the core) being accelerated by the outflow (see also
his Fig.13). Observations of absorption lines in diffuse clouds in front of bright quasars allow
\citet{Liszt_etal2001} to estimate a lower limit on the abundance ratio of [\ndhp]/[HCO$^+$]$< 0.002$ (at
2$\sigma$). \citet{Aikawa_etal2005} predict abundance ratios of [N$_2$H$^+$]/[HCO$^+$] within the range
$0.02-0.1$ for physical conditions relevant for pre-stellar collapsing dense cores. These chemical models
correspond to abundances of \ndhp{} and HCO$^+$, in agreement with observations of starless cores such as
L1544 \citep{Tafalla_etal2004,Caselli_etal2002b}. To first order, we get an averaged ratio [\ndhp]/[HCO$^+$]
on the order of $\sim 0.02-0.03$ (see Table \ref{tab:fit}). The abundance ratios we observe in the TTauri
disks are therefore similar to those observed and modelled in dense cores of cold molecular clouds.

\subsection{Reliability of chemical model}
Our non-mixing chemical model (described in Sect.~4.2) coupled to a 1+1D steady-state disk model with small
sub-micron size dust grains cannot reproduce the complexity of the disk structure. Surprisingly, we are able
to obtain reasonable global agreements with the observed column densities of \ndhp and HCO$^+$ around 400~AU
in all three disks (see Fig.~\ref{fig:coldens}). But we fail to reproduce in detail the observed slopes $p$.
The modelled \ndhp~ column density is in general too high in the outer disk ($R> 400$~AU), while the HCO$^+$
column density appears too low in the inner part of the disks. However, in the model as in the observations,
the [HCO$^+$]/[N$_2$H$^+$] ratio is almost constant with radius. More precisely, the value of $\sim 3\,
10^{11}$~cm$^{-2}$ for the column density of \ndhp in the DM~Tau disk is about 50\% higher than the observed
value, which is well within the error bars of a factor of 3 in the abundance due to the inaccuracy of the
reaction rates in the UMIST'95 ratefile \citep[see][]{Vasyunin_etal2004,Wakelam_etal2005}. The overall
agreement between the modelled and observed column densities of HCO$^+$ in DM~Tau, LkCa~15, and MWC~480 is
not as good as for \ndhp (see Fig.~\ref{fig:coldens}). The observed value of about $5\, 10^{12}$~cm$^{-2}$
that is nearly equal for the three disks is 3 times higher than predicted by the MWC~480 model and 3 times
lower than obtained with the DM~Tau and LkCa~15 disk models. Note that the computed HCO$^+$ column densities
also suffer from intrinsic uncertainties caused by the reaction rate uncertainties, about a factor of 3--4.

Since the chemistry of HCO$^+$ is sensitive to the grain growth, the presence of large grains, as usually
observed in disks with dust spectral index $\beta \simeq 1 -0.5$, may also account for the discrepancy
between the results of the applied chemical models and observational data, especially for HCO$^+$, but due to
the limitations of the current chemical prescription this cannot be easily checked.

\subsection{Determining the disk ionization fraction}
As shown in the previous section, the ionization fraction in different disk regions is dominated by distinct
ions, with C$^+$ the most abundant one in the upper layer. Is there any possibility of estimating the degree
of ionization in some disk regions or even in the entire disk?

The HCO$^+$ abundance gives a lower limit to the electron fraction, $[\mathrm{e}^-] > [$HCO$^+$]. The H$_2$
column density is uncertain: taking [$^{13}$CO] = 10$^{-7}$ and column density values from
\citet{Pietu_etal2007} leads to $[\mathrm{e}^-] > 2\,10^{-10}$. Following \citet[][~their Eq.6]{Qi_etal2003},
a lower limit to $[\mathrm{e}^-]$ can also be obtained from N$_2$H$^+$:
\begin{equation}
[\mathrm{e}^-]>{[\mathrm{CO}]}
\frac{k_2}{k_{rec}(\mathrm{HCO}^+)}\frac{[\mathrm{N}_2\mathrm{H}^+]}{[\mathrm{HCO}^+]} = 4.3\,10^{-4}
\frac{[\mathrm{N}_2\mathrm{H}^+]}{[\mathrm{HCO}^+]}.[\mathrm{CO}]
\end{equation}
where $k_2 = 8.8\,10^{-10}$ cm$^3$.s$^{-1}$ is the reaction rate coefficient of CO with N$_2$H$^+$ (CO +
N$_2$H$^+$ $\rightarrow$ HCO$^+$ + N$_2$) and $k_{rec}(\mathrm{HCO}^+)= 3\,
10^{-7}(T/300)^{-0.64}$~cm$^{3}$\,s$^{-1}$ is the dissociative recombination rate coefficient of HCO$^+$. The
numerical value is for $T = 15$~K. Thus N$_2$H$^+$ can be a more sensitive tracer of the ionization fraction
in disks than HCO$^+$ only if
\begin{equation}
\frac{[\mathrm{N}_2\mathrm{H}^+]}{[\mathrm{HCO}^+]} > \frac{1}{4.3\,10^{-4}} \frac{[\mathrm{HCO}^+]}
{[\mathrm{CO}]} .
\end{equation}
From \citet{Pietu_etal2007}, the [$^{13}$CO]/[HCO$^+$] ratio is $\simeq 600$. After correction of the
isotopologue column density ratio, around 20 when fractionation is accounted for \citep{Pietu_etal2007}, we
find that, if observed with a similar S/N, N$_2$H$^+$ can be a more sensitive tracer than HCO$^+$ only if
[N$_2$H$^+$]/[HCO$^+$] $ > 0.2$. With the measured abundance ratios, $\sim 0.02$, HCO$^+$ thus remains a
better tracer of the electron abundance than N$_2$H$^+$. Furthermore, these low abundance ratios no longer
support a high value for the gas phase [N$_2$]/[CO] ratio.

Another simple approach, but independent of the observations, to obtain an estimate of the disk fractional
ionization is to use the equation of chemical equilibrium \citep[e.g.,][]{Gammie_1996}:
\begin{equation}\label{gammie_eq}
[\mathrm{e}^-] = \sqrt{\frac{\zeta}{k_{\rm rec}n_{\rm H}}},
\end{equation}
where $\zeta$ is the ionization rate (in s$^{-1}$), $k_{\rm rec}$ is a typical recombination rate
(cm$^{3}$\,s$^{-1}$), and $n_{\rm H}$ is the hydrogen number density. \citet{Semenov_etal2004} have
demonstrated that the equilibrium approach is appropriate in most disk regions.  For the HCO$^+$ and \ndhp
ions, the recombination rates in our chemical network are $k_{\rm rec}=3\,
10^{-7}(T/300)^{-0.64}$~cm$^{3}$\,s$^{-1}$ and $k_{\rm rec}= 10^{-7}(T/300)^{-0.5}$~cm$^{3}$\,s$^{-1}$,
respectively \citep{Millar_etal1997,Geppert_etal2004,Geppert_etal2005}. The steady-state values of the
ionization degree in the warm molecular layers of the outer disk range are presented in
Table~\ref{gammie_res}. Temperatures were taken from Table~\ref{tab:disks}, the density from
\citet{Dartois_etal2003}, and the X-ray ionization rates from \citet{Glassgold_etal2005}. The cosmic ray rate
is $1.3\,10^{-17}$~s$^{-1}$. The values in Table~\ref{gammie_res} are consistent with the results of the
detailed chemical calculations shown in Fig.~\ref{fig:all_abunds} (first row, one scale height and radius
$\sim$ 200-500~AU). These values are high enough to allow magneto-rotational instability to operate in this
region of the disks \citep[see discussion in][]{Semenov_etal2004}.

\begin{table}\caption{Estimated ionization degrees}
\begin{center}
\begin{tabular}{lccccc}
\hline\hline
Source    & $T$     &  $n_{\rm H}$ & $\zeta_X$          & [e$^-$]          & [e$^-$] \\
          & (K)   &  cm$^{-3}$   & s$^{-1}$           & (HCO$^+$)        & (N$_2$H$^+$) \\
\hline
DM~Tau    & $15$  & $5\,10^6$ & $4\, 10^{-17}$  & $2\, 10^{-9}$ & $4\, 10^{-9}$ \\
LkCa~15   & $15$  & $5\,10^6$ & $10^{-16}$         & $3\, 10^{-9}$ & $7\, 10^{-9}$ \\
MWC~480   & $25$  & $5\,10^6$ & $1.3\, 10^{-17}$  & $2\, 10^{-9}$ & $3\, 10^{-9}$ \\
 \hline
\end{tabular}
\end{center} Estimates of the disk ionization degree in the intermediate layer derived from chemical
equilibrium approach (Eq.\ref{gammie_eq}), where the two values (col 5 and 6) depend on whether HCO$^+$ or
\ndhp\ dominates the recombination process. \label{gammie_res}
\end{table}

\section{Summary}

Using the IRAM array, we report new observations of \ndhp~ J=1-0 in three protoplanetary disks surrounding
two T Tauri stars: LkCa~15 and DM~Tau and the Herbig Ae/Be star MWC~480. We analyze these data by also taking
 the HCO$^+$ J=1-0 data from \citet{Pietu_etal2007} into account. For the three objects, we derive from the
observations the column densities and abundances and compare them to  model predictions obtained using a 1+1D
chemical code. Our main conclusions are the following:
\begin{itemize}

\item We detected \ndhp (at $5 \sigma$ level) in the disks surrounding LkCa~15 and DM~Tau. In the disk of
LkCa~15, the column density we observed is a factor $\sim 100$ below previous claims.

\item The \ndhp column densities derived from the observations imply an [\ndhp]/[HCO$^+$] ratio which is on
the order of the ratio found in cold dense cores $\sim 0.02-0.03$.

\item HCO$^+$ remains the more abundant molecular ion in disks. A lower limit of $2 \, 10^{-10}$ for the
ionization fraction can be obtained from the observations of HCO$^+$ and $^{13}$CO.

\item The overall column densities of N$_2$H$^+$ and HCO$^+$ and the [N$_2$H$^+$]/[HCO$^+$] ratio are
approximately reproduced by the chemical models, but their radial dependencies (slope $p$) are not. The
latter disagreement may be linked to the limits of the chemical modelling, which does not include processes
like grain growth or turbulent mixing. It could also be due to an over-simplistic disk structure, since
\citet{Pietu_etal2006} have shown that the dust distribution in LkCa~15 and MWC~480 are quite different from
the simple density distribution adopted here.
\end{itemize}
\begin{acknowledgements}
We acknowledge all the Plateau de Bure IRAM staff for their help during the observations. We also acknowledge
the Grenoble IRAM staff for their support during the first data reduction session of the CID project in
January 2006. We are thankful to Dmitry Wiebe for providing the disk physical models. The French Program of
Physico-Chemistry (PCMI) is thanked for providing funding for this project. D.S. and T.H. thank the Deutsche
Forschungsgemeinschaft, DFG project ``Research Group Laboratory Astrophysics'' (He 1935/17-2) for their
support.
\end{acknowledgements}

\bibliography{ms5385}

\begin{thebibliography}{49}
\expandafter\ifx\csname natexlab\endcsname\relax\def\natexlab#1{#1}\fi

\bibitem[{{Aikawa} {et~al.}(2005){Aikawa}, {Herbst}, {Roberts}, \&
  {Caselli}}]{Aikawa_etal2005}
{Aikawa}, Y., {Herbst}, E., {Roberts}, H., \& {Caselli}, P. 2005, \apj, 620,
  330

\bibitem[{{Aikawa} {et~al.}(2003){Aikawa}, {Momose}, {Thi}, {van Zadelhoff},
  {Qi}, {Blake}, \& {van Dishoeck}}]{Aikawa_etal2003}
{Aikawa}, Y., {Momose}, M., {Thi}, W.-F., {et~al.} 2003, \pasj, 55, 11

\bibitem[{{Anicich} \& {Huntress}(1986)}]{Anicich_Huntress86}
{Anicich}, V.~G. \& {Huntress}, W.~T. 1986, \apjs, 62, 553

\bibitem[{{Beckwith} \& {Sargent}(1993)}]{Beckwith_Sargent_1993}
{Beckwith}, S.~V.~W. \& {Sargent}, A.~I. 1993, \apj, 402, 280

\bibitem[{{Bergin} {et~al.}(2004){Bergin}, {Calvet}, {Sitko}, {Abgrall},
  {D'Alessio}, {Herczeg}, {Roueff}, {Qi}, {Lynch}, {Russell}, {Brafford}, \&
  {Perry}}]{Bergin_etal2004}
{Bergin}, E., {Calvet}, N., {Sitko}, M.~L., {et~al.} 2004, \apjl, 614, L133

\bibitem[{{Bergin} {et~al.}(2006){Bergin}, {Aikawa}, {Blake}, \& {van
  Dischoeck}}]{Bergin_etal2006}
{Bergin}, E.~A., {Aikawa}, Y., {Blake}, G.~A., \& {van Dischoeck}, E.~F. 2006,
  in Protostars and Planets V, Proceedings of the Conference held October
  24-28, 2005, in Hilton Waikoloa Village, Hawai'i., --+

\bibitem[{{Bisschop} {et~al.}(2006){Bisschop}, {Fraser}, {{\"O}berg}, {van
  Dishoeck}, \& {Schlemmer}}]{Bisschop_etal2006}
{Bisschop}, S.~E., {Fraser}, H.~J., {{\"O}berg}, K.~I., {van Dishoeck}, E.~F.,
  \& {Schlemmer}, S. 2006, \aap, 449, 1297

\bibitem[{{Calvet} {et~al.}(2005){Calvet}, {D'Alessio}, {Watson},
  {Franco-Hern{\'a}ndez}, {Furlan}, {Green}, {Sutter}, {Forrest}, {Hartmann},
  {Uchida}, {Keller}, {Sargent}, {Najita}, {Herter}, {Barry}, \&
  {Hall}}]{Calvet_etal2005}
{Calvet}, N., {D'Alessio}, P., {Watson}, D.~M., {et~al.} 2005, \apjl, 630, L185

\bibitem[{{Caselli} {et~al.}(2002{\natexlab{a}}){Caselli}, {Walmsley},
  {Zucconi}, {Tafalla}, {Dore}, \& {Myers}}]{Caselli_etal2002a}
{Caselli}, P., {Walmsley}, C.~M., {Zucconi}, A., {et~al.} 2002{\natexlab{a}},
  \apj, 565, 331

\bibitem[{{Caselli} {et~al.}(2002{\natexlab{b}}){Caselli}, {Walmsley},
  {Zucconi}, {Tafalla}, {Dore}, \& {Myers}}]{Caselli_etal2002b}
{Caselli}, P., {Walmsley}, C.~M., {Zucconi}, A., {et~al.} 2002{\natexlab{b}},
  \apj, 565, 344

\bibitem[{{Ceccarelli} \& {Dominik}(2005)}]{Ceccarelli_Dominik2005}
{Ceccarelli}, C. \& {Dominik}, C. 2005, \aap, 440, 583

\bibitem[{{Crapsi} {et~al.}(2005){Crapsi}, {Devries}, {Huard}, {Lee}, {Myers},
  {Ridge}, {Bourke}, {Evans}, {J{\o}rgensen}, {Kauffmann}, {Lee}, {Shirley}, \&
  {Young}}]{Crapsi_etal2005}
{Crapsi}, A., {Devries}, C.~H., {Huard}, T.~L., {et~al.} 2005, \aap, 439, 1023

\bibitem[{{D'Alessio} {et~al.}(1999){D'Alessio}, {Calvet}, {Hartmann},
  {Lizano}, \& {Cant{\'o}}}]{DAlessio_etal1999}
{D'Alessio}, P., {Calvet}, N., {Hartmann}, L., {Lizano}, S., \& {Cant{\'o}}, J.
  1999, \apj, 527, 893

\bibitem[{{Dartois} {et~al.}(2003){Dartois}, {Dutrey}, \&
  {Guilloteau}}]{Dartois_etal2003}
{Dartois}, E., {Dutrey}, A., \& {Guilloteau}, S. 2003, \aap, 399, 773

\bibitem[{{Draine}(1978)}]{Draine1978}
{Draine}, B.~T. 1978, \apjs, 36, 595

\bibitem[{{Dutrey} {et~al.}(1997){Dutrey}, {Guilloteau}, \&
  {Guelin}}]{Dutrey_etal1997}
{Dutrey}, A., {Guilloteau}, S., \& {Guelin}, M. 1997, \aap, 317, L55

\bibitem[{{Gammie}(1996)}]{Gammie_1996}
{Gammie}, C.~F. 1996, \apj, 457, 355

\bibitem[{{Geppert} {et~al.}(2004){Geppert}, {Thomas}, {Semaniak}, {Ehlerding},
  {Millar}, {{\"O}sterdahl}, {af Ugglas}, {Djuri{\'c}}, {Pa{\'a}l}, \&
  {Larsson}}]{Geppert_etal2004}
{Geppert}, W.~D., {Thomas}, R., {Semaniak}, J., {et~al.} 2004, \apj, 609, 459

\bibitem[{{Geppert} {et~al.}(2005){Geppert}, {Thomas}, {Ehlerding}, {Hellberg},
  {{\"O}sterdahl}, {Hamberg}, {Semaniak}, {Zhaunerchyk}, {Kaminska},
  {K{\"a}llberg}, {Paal}, \& {Larsson}}]{Geppert_etal2005}
{Geppert}, W.~D., {Thomas}, R.~D., {Ehlerding}, A., {et~al.} 2005, Journal of
  Physics Conference Series, 4, 26

\bibitem[{{Glassgold} {et~al.}(2005){Glassgold}, {Feigelson}, {Montmerle}, \&
  {Wolk}}]{Glassgold_etal2005}
{Glassgold}, A.~E., {Feigelson}, E.~D., {Montmerle}, T., \& {Wolk}, S. 2005, in
  ASP Conf. Ser. 341: Chondrites and the Protoplanetary Disk, 165--+

\bibitem[{{Glassgold} {et~al.}(1997{\natexlab{a}}){Glassgold}, {Najita}, \&
  {Igea}}]{GNI97a}
{Glassgold}, A.~E., {Najita}, J., \& {Igea}, J. 1997{\natexlab{a}}, \apj, 480,
  344

\bibitem[{{Glassgold} {et~al.}(1997{\natexlab{b}}){Glassgold}, {Najita}, \&
  {Igea}}]{GNI97b}
{Glassgold}, A.~E., {Najita}, J., \& {Igea}, J. 1997{\natexlab{b}}, \apj, 485,
  920

\bibitem[{{Guilloteau} \& {Dutrey}(1998)}]{Guilloteau_Dutrey1998}
{Guilloteau}, S. \& {Dutrey}, A. 1998, \aap, 339, 467

\bibitem[{{Guilloteau} {et~al.}(2006){Guilloteau}, {Pi{\'e}tu}, {Dutrey}, \&
  {Gu{\'e}lin}}]{Guilloteau_etal2006}
{Guilloteau}, S., {Pi{\'e}tu}, V., {Dutrey}, A., \& {Gu{\'e}lin}, M. 2006,
  \aap, 448, L5

\bibitem[{{Hasegawa} \& {Herbst}(1993)}]{Hasegawa_etal1993}
{Hasegawa}, T.~I. \& {Herbst}, E. 1993, \mnras, 263, 589

\bibitem[{{Hasegawa} {et~al.}(1992){Hasegawa}, {Herbst}, \&
  {Leung}}]{Hasegawa_etal1992}
{Hasegawa}, T.~I., {Herbst}, E., \& {Leung}, C.~M. 1992, \apjs, 82, 167

\bibitem[{{J{\o}rgensen}(2004)}]{Jorgensen_2004}
{J{\o}rgensen}, J.~K. 2004, \aap, 424, 589

\bibitem[{{Kamp} {et~al.}(2003){Kamp}, {van Zadelhoff}, {van Dishoeck}, \&
  {Stark}}]{Kamp_etal2003}
{Kamp}, I., {van Zadelhoff}, G.-J., {van Dishoeck}, E.~F., \& {Stark}, R. 2003,
  \aap, 397, 1129

\bibitem[{{Kastner} {et~al.}(1997){Kastner}, {Zuckerman}, {Weintraub}, \&
  {Forveille}}]{Kastner_etal1997}
{Kastner}, J.~H., {Zuckerman}, B., {Weintraub}, D.~A., \& {Forveille}, T. 1997,
  Science, 277, 67

\bibitem[{{Kurucz}(1993)}]{Kurucz1993}
{Kurucz}, R. 1993, ATLAS9 Stellar Atmosphere Programs and 2 km/s grid.~Kurucz
  CD-ROM No.~13.~ Cambridge, Mass.: Smithsonian Astrophysical Observatory,
  1993., 13

\bibitem[{{Liszt} \& {Lucas}(2001)}]{Liszt_etal2001}
{Liszt}, H. \& {Lucas}, R. 2001, \aap, 370, 576

\bibitem[{{Mannings} {et~al.}(1997){Mannings}, {Koerner}, \&
  {Sargent}}]{Mannings_etal1997}
{Mannings}, V., {Koerner}, D.~W., \& {Sargent}, A.~I. 1997, \nat, 388, 555

\bibitem[{{Millar} {et~al.}(1997){Millar}, {Farquhar}, \&
  {Willacy}}]{Millar_etal1997}
{Millar}, T.~J., {Farquhar}, P.~R.~A., \& {Willacy}, K. 1997, \aaps, 121, 139

\bibitem[{{Pagani} {et~al.}(2005){Pagani}, {Pardo}, {Apponi}, {Bacmann}, \&
  {Cabrit}}]{Pagani_etal2005}
{Pagani}, L., {Pardo}, J.-R., {Apponi}, A.~J., {Bacmann}, A., \& {Cabrit}, S.
  2005, \aap, 429, 181

\bibitem[{{Pi\'etu} {et~al.}(2007){Pi\'etu}, {Dutrey}, \&
  {Guilloteau}}]{Pietu_etal2007}
{Pi\'etu}, V., {Dutrey}, A., \& {Guilloteau}, S. 2007, \aap, submitted

\bibitem[{{Pi{\'e}tu} {et~al.}(2006){Pi{\'e}tu}, {Dutrey}, {Guilloteau},
  {Chapillon}, \& {Pety}}]{Pietu_etal2006}
{Pi{\'e}tu}, V., {Dutrey}, A., {Guilloteau}, S., {Chapillon}, E., \& {Pety}, J.
  2006, \aap, 460, L43

\bibitem[{{Qi} {et~al.}(2004){Qi}, {Ho}, {Wilner}, {Takakuwa}, {Hirano},
  {Ohashi}, {Bourke}, {Zhang}, {Blake}, {Hogerheijde}, {Saito}, {Choi}, \&
  {Yang}}]{Qi_etal2004}
{Qi}, C., {Ho}, P.~T.~P., {Wilner}, D.~J., {et~al.} 2004, \apjl, 616, L11

\bibitem[{{Qi} {et~al.}(2003){Qi}, {Kessler}, {Koerner}, {Sargent}, \&
  {Blake}}]{Qi_etal2003}
{Qi}, C., {Kessler}, J.~E., {Koerner}, D.~W., {Sargent}, A.~I., \& {Blake},
  G.~A. 2003, \apj, 597, 986

\bibitem[{{Semenov} {et~al.}(2005){Semenov}, {Pavlyuchenkov}, {Schreyer},
  {Henning}, {Dullemond}, \& {Bacmann}}]{Semenov_etal2005}
{Semenov}, D., {Pavlyuchenkov}, Y., {Schreyer}, K., {et~al.} 2005, \apj, 621,
  853

\bibitem[{{Semenov} {et~al.}(2004){Semenov}, {Wiebe}, \&
  {Henning}}]{Semenov_etal2004}
{Semenov}, D., {Wiebe}, D., \& {Henning}, T. 2004, \aap, 417, 93

\bibitem[{{Semenov} {et~al.}(2006){Semenov}, {Wiebe}, \&
  {Henning}}]{Semenov_etal2006}
{Semenov}, D., {Wiebe}, D., \& {Henning}, T. 2006, \apjl, 647, L57

\bibitem[{{Simon} {et~al.}(2000){Simon}, {Dutrey}, \&
  {Guilloteau}}]{Simon_etal2000}
{Simon}, M., {Dutrey}, A., \& {Guilloteau}, S. 2000, \apj, 545, 1034

\bibitem[{{Tafalla} {et~al.}(2004){Tafalla}, {Myers}, {Caselli}, \&
  {Walmsley}}]{Tafalla_etal2004}
{Tafalla}, M., {Myers}, P.~C., {Caselli}, P., \& {Walmsley}, C.~M. 2004, \aap,
  416, 191

\bibitem[{{Thi} {et~al.}(2004){Thi}, {van Zadelhoff}, \& {van
  Dishoeck}}]{Thi_etal2004}
{Thi}, W.-F., {van Zadelhoff}, G.-J., \& {van Dishoeck}, E.~F. 2004, \aap, 425,
  955

\bibitem[{{van Zadelhoff} {et~al.}(2003){van Zadelhoff}, {Aikawa},
  {Hogerheijde}, \& {van Dishoeck}}]{vanZadelhoff_etal2003}
{van Zadelhoff}, G.-J., {Aikawa}, Y., {Hogerheijde}, M.~R., \& {van Dishoeck},
  E.~F. 2003, \aap, 397, 789

\bibitem[{{van Zadelhoff} {et~al.}(2001){van Zadelhoff}, {van Dishoeck}, {Thi},
  \& {Blake}}]{Zadelhoff_etal2001}
{van Zadelhoff}, G.-J., {van Dishoeck}, E.~F., {Thi}, W.-F., \& {Blake}, G.~A.
  2001, \aap, 377, 566

\bibitem[{{Vasyunin} {et~al.}(2006){Vasyunin}, {Semenov}, {Sobolev}, \&
  {Henning}}]{Vasyunin_etal2006}
{Vasyunin}, A.~I., {Semenov}, D.~A., {Sobolev}, A.~M., \& {Henning}, T. 2006,
  \apj, in preparation

\bibitem[{{Vasyunin} {et~al.}(2004){Vasyunin}, {Sobolev}, {Wiebe}, \&
  {Semenov}}]{Vasyunin_etal2004}
{Vasyunin}, A.~I., {Sobolev}, A.~M., {Wiebe}, D.~S., \& {Semenov}, D.~A. 2004,
  Astronomy Letters, 30, 566

\bibitem[{{Wakelam} {et~al.}(2005){Wakelam}, {Selsis}, {Herbst}, \&
  {Caselli}}]{Wakelam_etal2005}
{Wakelam}, V., {Selsis}, F., {Herbst}, E., \& {Caselli}, P. 2005, \aap, 444,
  883

\end{thebibliography}
\bibliographystyle{aa}
\end{document}